\documentclass[12pt,titlepage,nofootinbib,superscriptaddress,showkeys]{revtex4-1}

\usepackage{graphicx}
\usepackage{psfrag}

\usepackage{amsmath,amssymb,mathrsfs}
\usepackage[bookmarks=false]{hyperref} 
\hypersetup{pdfstartview=FitH,pdfhighlight=/O,colorlinks=false}

\usepackage{cases}
\usepackage{subfigure}

\usepackage[left=1.2in,right=1.2in,top=1.5in,bottom=1.5in]{geometry} 
\usepackage{enumitem}

\begin{document}
%%%%%%%%%%%%%%%%%%%%%%%%%%%%%%%%%%%%%%%%%%%%%%%%%%%

\title{\large \bf Entanglement time in the primordial universe}

\author{Eugenio Bianchi}\email{ebianchi@gravity.psu.edu}
\author{Lucas Hackl}\email{lucas.hackl@psu.edu}
\author{Nelson Yokomizo}\email{yokomizo@gravity.psu.edu}
\affiliation{Institute for Gravitation and the Cosmos, Physics Department,
Penn State, University Park, PA 16802, USA\\
${}$}

%\date{\small\today}

\begin{abstract}
\noindent {\bf Abstract}. We investigate the behavior of the entanglement entropy of space in the primordial phase of the universe before the beginning of cosmic inflation. We argue that in this phase the entanglement entropy of a region of space grows from a zero-law to an area-law. This behavior provides a quantum version of the classical BKL conjecture that spatially separated points decouple in the approach to a cosmological singularity. We show that the relational growth of the entanglement entropy with the scale factor provides a new statistical notion of arrow of time in quantum gravity. The growth of entanglement in the pre-inflationary phase provides a mechanism for the production of the quantum correlations present at the beginning of inflation and imprinted in the CMB sky.\footnote{Honorable Mention in the Gravity Research Foundation 2015 Essay Competition.\\[-1em]}
%\\[2em]
%\begin{center}
%%{\bf Essay written for the Gravity Research Foundation\\ 2015 Awards for Essays on Gravitation}
%\end{center}
\end{abstract}

\maketitle

\newpage

\section{Introduction}

Current observations of the cosmic microwave radiation (CMB) \cite{Ade:2015lrj}, together with the inflationary paradigm, indicate that at the beginning of cosmic inflation the universe was in a pure state with highly-correlated quantum fluctuations. These correlated quantum fluctuations are imprinted in the CMB sky and correspond to an \emph{area law} \cite{sorkin1983entropy,bombelli1986quantum,srednicki1993entropy,eisert2010colloquium}
 for the entanglement entropy of quantum fields.

In this essay we explore the behavior of entanglement before the beginning of inflation, that is in the primordial universe. In classical general relativity, the Belinsky-Khalatnikov-Lifshitz (BKL) conjecture \cite{Belinsky:1970ew} indicates that the spatial coupling of degrees of freedom is suppressed in the approach to a space-like singularity. In the quantum theory this phenomenon corresponds to the suppression of correlations at space-like separation, i.e. a suppression of the entanglement entropy. We consider a class of candidate solutions of the Wheeler-deWitt (WdW) equation that have exactly this behavior: a vanishing entanglement entropy in the limit of small scale factor. We argue that the dynamics entangles the quantum degrees of freedom of space resulting in a growth of the entanglement entropy with the scale factor: the entanglement entropy grows -- following a quantum version of the second law of thermodynamics -- until it saturates to an equilibrium state where the area law holds and standard quantum field theory on a classical background geometry applies.

\section{The candle and the pendulum}

Our analysis brings together two independent conceptual insights regarding the nature of time: relational time and the thermodynamic arrow of time.

The notion of \emph{relational time} is relevant for generally-covariant systems where the parameter time $t$ is not an observable \cite{rovelli1991time}. What a generally-covariant  theory predicts and experiments can test is the value of an observable $\mathcal{O}$ conditioned to having measured the value of  another observable $a$. Introducing  the parameter time $t$, we have the partial observables $a(t)$ and  $\mathcal{O}(t)$ and the prediction  $\mathcal{O}[a]\equiv\mathcal{O}(t(a))$ of finding the value $\mathcal{O}$ given the value $a$. For instance, we can describe the correlation of the position of a pendulum with the position of a ball rolling down an inclined plane. This timeless description of the dynamics applies to classical systems as well as to quantum systems as discussed in Sec.~\ref{sec:time-lessQM}.

The notion of \emph{thermodynamic arrow of time} arises in isolated macroscopic systems where, according to the second law of thermodynamics, the entropy does not decrease in time. The microscopic foundations of this law rest on Boltzmann's statistical explanation of the observed irreversible behavior of macroscopic systems \cite{lebowitz1993boltzmann,brush1966kinetic}: the reversible microscopic dynamics of the system typically results in the evolution from low statistical entropy towards the maximum allowed entropy, the equilibrium state. The argument involves two ingredients: a choice of coarse graining of microscopic degrees of freedom and the preparation of the system in an initial microscopic configuration with low statistical entropy. For instance, a burning candle is initially in a low-entropy state and the height of the candle can be used to measure a time lapse and its direction. Fixing the initial conditions is what breaks the time-reversal symmetry.

It can be argued that to measure the passing of time we need both a pendulum and a candle. Our best clocks have oscillators with all frictions eliminated as far as possible. A perfect oscillator is one that perfectly conceals the arrow of time, while accurately measuring the length of a given interval of time, proportional to the number of oscillations. Correlations of the position of the pendulum and the height of a burning candle then establish the direction of the flow of time. Both instruments are required if one wants to measure a directed interval of time. In fact, mechanical watches have this system built in: a good oscillator and an escapement mechanism that is initially in a low-entropy state which introduces a small but necessary dissipation in order to keep track of the number of oscillations, making the hands advance in one direction.

We argue that the primordial universe is no different: the entanglement entropy of a region of space typically grows with the scale factor resulting in an entanglement arrow of time in an isolated and timeless quantum system.

In the next two sections we illustrate the notion of relational time in time-less quantum mechanics (Sec.~\ref{sec:time-lessQM}) and the second law of thermodynamics of quantum systems initially prepared in a pure un-entangled state (Sec.~\ref{sec:entanglement-2nd law}). The application to quantum gravity and its relevance to the description of the pre-inflationary universe is discussed in Sec.~\ref{sec:WdW}.

\section{Relational time in time-less quantum mechanics}
\label{sec:time-lessQM}

Differently from the Schr\"odinger equation, the Wheeler-deWitt equation in quantum gravity does not contain a parameter time $t$. An ordinary quantum mechanical system showing this same behavior is a stationary state, i.e. an eigenstate of the energy \cite{Rovelli:2004tv}. As an example, consider two uncoupled harmonic oscillators with Hamiltonians $\hat{H}_1=\hbar \omega_1 (a_1^\dagger a_1+\frac{1}{2})$ and $\hat{H}_2=\hbar \omega_2 (a_2^\dagger a_2+\frac{1}{2})$. The level of energy $E$ of the combined system has finite degeneracy if the frequencies are proportional. We take $\omega_2=M \omega_1$ and $E=\hbar\omega_1(N+\frac{M+1}{2})$, with $N$ a multiple of the integer $M$.
The WdW equation for this simple system reads
\begin{equation}
\hat{H}\Psi(x_1,x_2)=0\,.
\label{eq:WdW}
\end{equation}
where $\hat{H}=\hat{H}_1+\hat{H}_2-E$, and the most general solution $|s\rangle$ has the form
\begin{equation}
\Psi_s(x_1,x_2)=\sum_{n=0}^{N/M}c_n\;\, \psi_{N-Mn}(x_1)\,\psi_{n}(x_2)
\label{eq:Psis}
\end{equation}
where $\psi_{n_1}(x_1)$ and $\psi_{n_2}(x_2)$ are eigenstates of each of the two oscillators. While there is no time in this quantum system, we can still speak about the evolution of the expectation value of an observable conditioned to a value of the other. In the following we discuss the evolution of the expectation value of the position $\langle x_2 \rangle$ of the second oscillator, conditioned to the value $x_1=a$ of the position of the first oscillator. See also \cite{wootters1984time,rovelli1991time,page1994clock,Rovelli:2001bz,gambini2009conditional,moreva2014time}.

\begin{figure}[t]
\includegraphics[width = 30em]{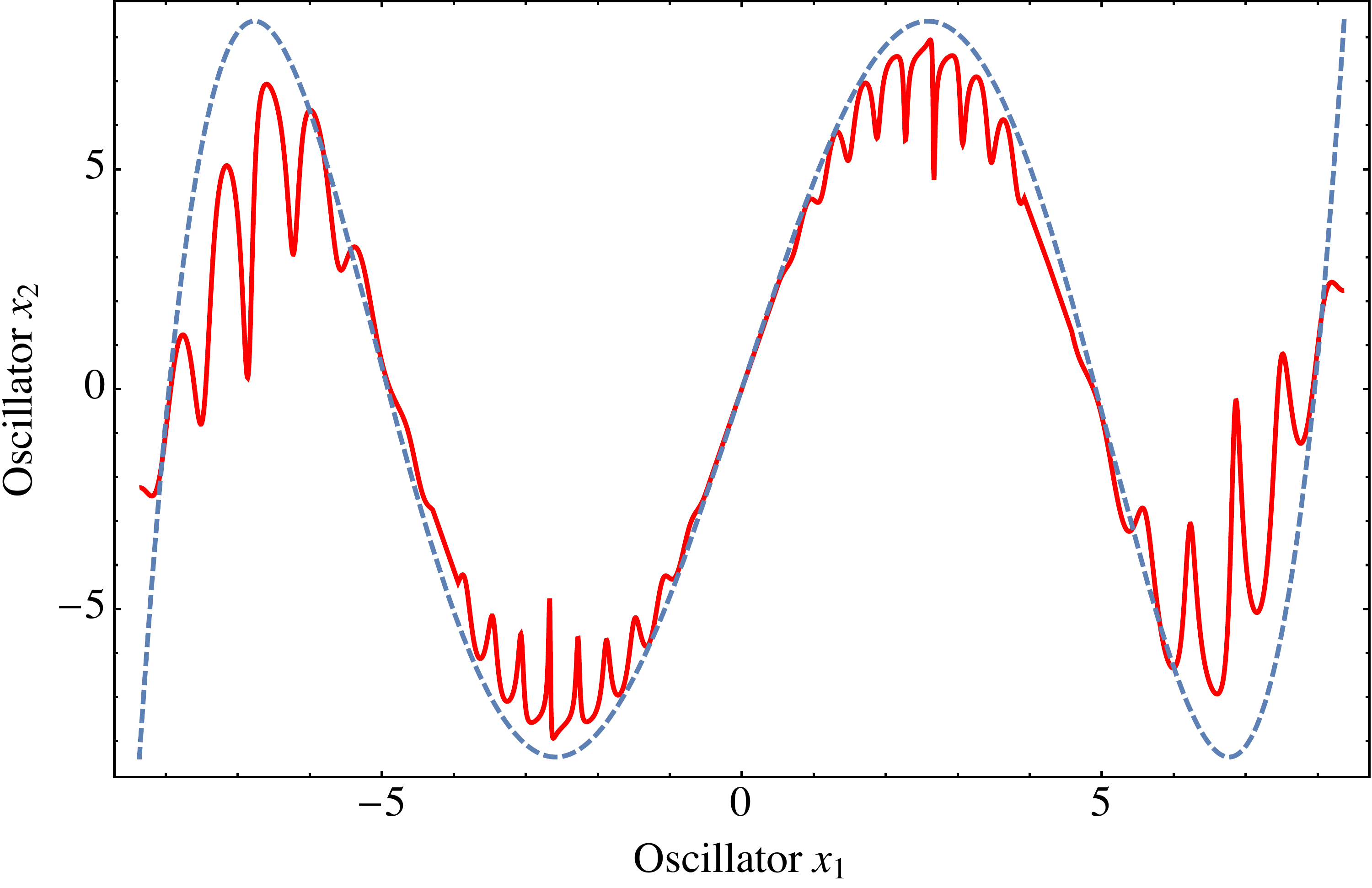}
\caption{Classical and quantum relational observables for two uncoupled harmonic oscillators in a state of definite total energy: The dashed curve shows the classical position $x_2$ of the second oscillator conditioned to the first oscillator having at position $x_1$. The continuous curve shows its quantum version, the expectation value of $x_2$ conditioned to a measurement of the first oscillator.}
\label{fig:timeless}
\end{figure}

The setting is the standard one of positive-operator valued measures (POVM). We consider a Gaussian POVM in terms of squeezed coherent states $|\alpha\rangle$ of the first oscillator, ($\alpha\in \mathbb{C}$). The completeness relation reads $1\!\!1=\int d^2\alpha\, F_\alpha F_\alpha^\dagger\;$ with $F_\alpha=\frac{1}{\sqrt{\pi}}|\alpha\rangle\langle \alpha|\otimes 1\!\!1_2$. The state of the second oscillator conditioned to the value $\alpha$ for the first is
\begin{equation}
\rho_2(\alpha)=\frac{\text{Tr}_1\Big(F_\alpha |s\rangle\langle s|F_\alpha^\dagger\Big)}{\text{Tr}_1\text{Tr}_2\Big(F_\alpha |s\rangle\langle s|F_\alpha^\dagger\Big)}
\label{eq:}
\end{equation}
where $\text{Tr}_1$ and $\text{Tr}_2$ are the traces in the Hilbert space of each  oscillator. The conditioned value of the position of the second oscillator is $x_2(\alpha)=\text{Tr}_2\big(\hat{x}_2\, \rho_2(\alpha)\big)$.
%\begin{equation}
%x_2(\alpha)=\text{Tr}_2\Big(\hat{x}_2\, \rho_2(\alpha)\Big).
%\label{eq:}
%\end{equation}
Conditioning the first oscillator to a definite position $x_1=a$ corresponds to the limit of infinite squeezing of the coherent state $|\alpha\rangle$, i.e. the limit of a Gaussian to a Dirac delta function centered at $x_1=a$. In this case we obtain a rather simple formula for the conditioned state:
\begin{equation}
\rho_2(a;x_2,x_2')=\frac{\Psi_s(a,x_2)^*\,\Psi_s(a,x'_2)}{\int dx''_2\, |\Psi_s(a,x''_2)|^2}\,.
\label{eq:}
\end{equation}
Notice that this reduced state is pure: the entanglement between the two oscillators has been exploited in obtaining the conditioned state. The conditioned position $x_2(a)$  of the second oscillator is easily computed.
%\begin{equation}
%x_2(a)=\frac{\int dx'_2\;x'_2\, \Psi_s(a,x'_2)|^2}{\int dx''_2\, |\Psi_s(a,x''_2)|^2}\,.
%\label{eq:}
%\end{equation}
Fig. (\ref{fig:timeless}) shows a plot of the evolution of the position $x_2$ in the relational time $x_1=a$  for a constrained coherent state of the form (\ref{eq:Psis}).
% with components
%\begin{equation}
%c_n=\frac{(N-M\bar{n})^{\frac{N-Mn}{2}}\;\,\bar{n}^{\frac{n}{2}}}{\sqrt{(N-Mn)!\,n!}}\,.
%\label{eq:}
%\end{equation}
In the plot the quantum relational evolution (continuous line) is compared to its classical counterpart (dashed line).

\section{Entanglement entropy and the second law}
\label{sec:entanglement-2nd law}

\begin{figure}[t]
\includegraphics[width = 30em]{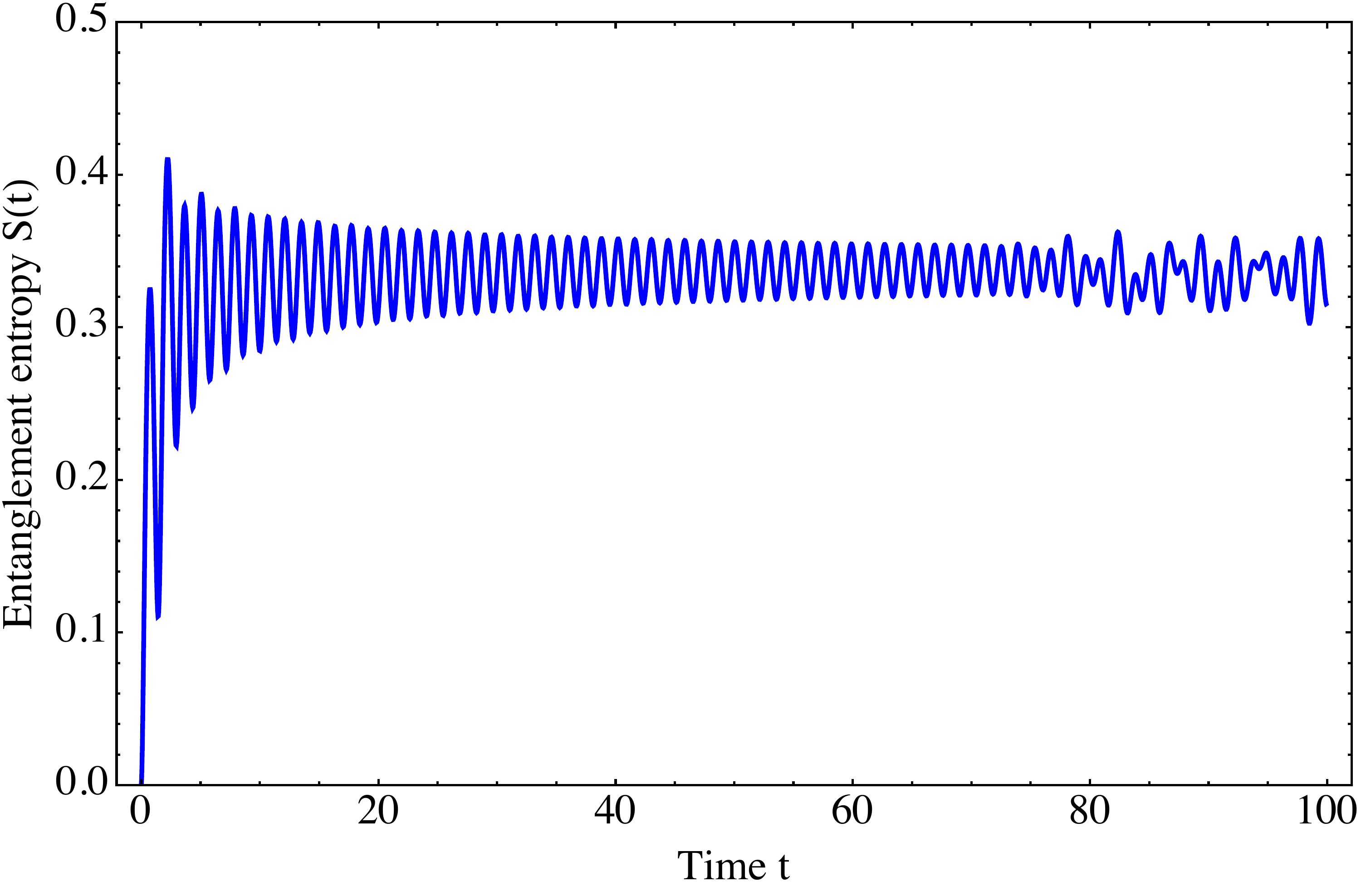}
\caption{Entanglement production in a 1d lattice with 99 oscillators: The plot shows the growing and equilibration of the entanglement entropy of a single oscillator.}
\label{fig:entropyprod}
\end{figure}

Boltzmann's notion of thermodynamic arrow of time, originally proposed for isolated classical systems, applies to quantum systems in a pure state as well. The coarse graining of the microscopic degrees of freedom is given by a choice of sub-algebra of observables $R$ corresponding to a splitting of the Hilbert space of the system in a tensor product $\mathcal{H}=\mathcal{H}_R\otimes \mathcal{H}_{\bar{R}}$. A pure state $|s\rangle$ restricted to this sub-algebra results in a density matrix $\rho_R=\text{Tr}_{\bar{R}}(|s\rangle\langle s|)$ that is \emph{typically} mixed because of entanglement between $R$ and $\bar{R}$. The \emph{entanglement entropy} is the entropy of measurements of the observables in $R$ on the state $|s\rangle$ and is defined as the von Neumann entropy of the restricted state $S_R(|s\rangle)=-\text{Tr}_R  (\rho_R \log \rho_R)$.  While the Schr\"odinger equation is invariant under time reversal, the evolution of a system initially prepared in a state of low entanglement entropy typically results in an increase of the entanglement entropy towards its value in the equilibrium configuration. This mechanism provides the foundations of the second law of thermodynamics in isolated quantum systems \cite{linden2009quantum} and is illustrated below in a toy model.

The toy model consists of a discretized free scalar quantum field. The system is equivalent to a collection of harmonic oscillators, one at every lattice site and coupled through the discretized Laplacian on the lattice. In particular, the tensor product of the individual oscillator vacua $|I\rangle=|0\rangle\otimes\cdots\otimes|0\rangle$ has zero entanglement. If we use $|I\rangle$ as initial state for the time evolution through the field theory's Hamiltonian, the entanglement entropy will start growing from $S=0$ with the asymptotics of $S\sim t^2\log(1/t)$ and finally approach its equilibrium value in an oscillating behavior. Recurrence occurs on a much longer time scale. In fig. \ref{fig:entropyprod}, we show how this saturation takes place on a 1d lattice.

\section{WdW equation, BKL behavior and entanglement time}
\label{sec:WdW}

Consider the WdW equation for gravity and an inflaton scalar field
\begin{equation}
\hat{H}\,\Psi[g_{ij}(x),\varphi(x)]=0\,.
\label{eq:}
\end{equation}
The Hamiltonian constraint $\hat{H}=T+U$ consists of a kinetic term $T$ that involves the momenta $\pi^{ij}(x)$ and $\pi_\varphi(x)$ and a potential term $U$ that involves the spatial metric $h_{ij}(x)$, the scalar field $\varphi(x)$ \emph{and} their spatial derivatives.\footnote{The potential $U$ includes both the spatial curvature ${}^{(3)}\!R$ of the metric and the spatial derivatives $g^{ij}\partial_i \varphi \partial_j \varphi$ of the field, as well as the inflaton potential $V(\varphi)$.} While the kinetic term is ultra-local (in the sense that it does not contain spatial derivatives), the potential term is only local and couples nearby points. For small perturbations of a flat homogeneous and isotropic configuration we can write $h_{ij}(x)=a^2\,(\delta_{ij} +\epsilon_{ij}(x))$ and $\varphi(x)=\phi+\delta \varphi(x)$ where $a$ is the scale factor and $\phi$ the average value of the inflaton field. Correspondingly the wavefunction of the universe reads $\Psi[a, \phi, \epsilon_{ij}(x),\delta \varphi(x)]$. In the approach to a space-like cosmological singularity where the scale factor vanishes, $a\to 0$,  the kinetic term $T$ can formally be shown to be dominating over the potential term $U$. As the potential $U$ is the one that couples nearby points in the WdW equation, neglecting it altogether allows us to easily find solutions: they are simply a product over points of solutions at each point. When $U$ is not neglected this behavior can be imposed as a boundary condition:
\begin{equation}
\lim_{a\to 0}\Psi[a, \phi, \epsilon_{ij}(x),\delta \varphi(x)]=\prod_{\vec{x}}\psi\big(\phi, \epsilon_{ij}(x),\delta \varphi(x)\big)\,.
\label{eq:QuantumBKL}
\end{equation}
In a candidate full theory of quantum gravity, Eq.(\ref{eq:QuantumBKL}) serves as a proposed quantum version of the classical BKL conjecture about the behavior of the metric and the matter fields in the approach to a space-like singularity: \emph{the entanglement entropy between a region of space and its complement vanishes in the limit of Planck-scale curvature}.

\begin{figure}[t]
\includegraphics[width = 30em]{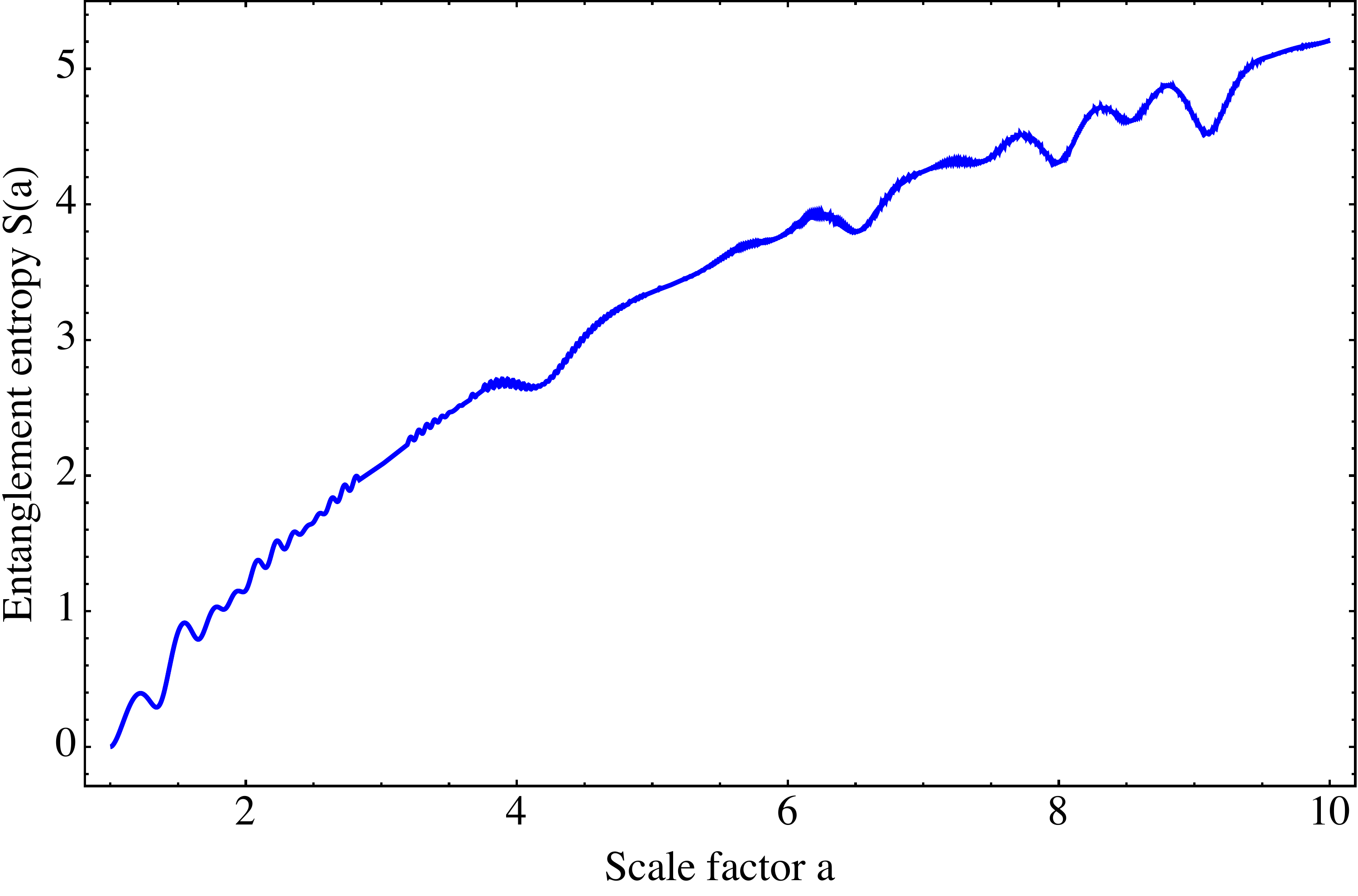}
\caption{Relational entanglement entropy in the primordial universe: This plot shows the entanglement entropy of Gaussian inflaton fluctuation as a function of the primordial scale factor $a$ provided that the field was prepared in a completely un-entangled initial state at $a_0=1$. The inflaton field is modeled as a scalar field on a finite 3d lattice.}
\label{fig:WdW}
\end{figure}

As discussed in Sec.~\ref{sec:entanglement-2nd law}, the time evolution of an initially un-entangled state typically results in a growth of the entanglement entropy. In the present case, Eq.~\ref{eq:QuantumBKL} provides the initially un-entangled state. On the other hand the discussion of Sec.~\ref{sec:entanglement-2nd law} presupposes the existence of an external time $t$ in which the entropy grows, an ingredient that is missing in the WdW equation.
Applying the logic discussed in Sec.~\ref{sec:time-lessQM} for time-less quantum systems, we can ask relational questions about the state. The question we are interested in is: given the state $\Psi[a, \phi, \epsilon_{ij}(x),\delta \varphi(x)]$, what is the entanglement entropy of a region of space $R$ conditioned to the observation of a given value $a$ of the scale factor? The strategy is to compute the state $\rho_R(a)$ restricted to the region $R$ and conditioned to the value $a$ of the scale factor. The entanglement entropy as a function of the scale factor is then given by
\begin{equation}
S_R(a)=-\text{Tr}_R\Big(\rho_R(a) \log \rho_R(a)\Big)\,.
\label{eq:SR}
\end{equation}
We investigated numerically the behavior of $S_R(a)$ for a lattice version of the inflaton field fluctuations $\delta \varphi(x)$, retaining only quadratic terms in the fluctuation \cite{Bianchi:2015fra}. The results are reported in Fig.~\ref{fig:WdW}: the entanglement entropy grows with the scale factor therefore providing a statistical arrow of time in the primordial universe, the \emph{entanglement time}.

The standard inflationary paradigm posits that, before the hot big bang phase, the universe went through a phase of exponential expansion with the linear perturbations of the gravitational and the inflaton fields  \emph{prepared} in a pure state with short-distance correlations matching the flat Minkowski vacuum ones, the Bunch-Davies vacuum. This is a state with vanishing expectation value of the Weyl curvature and small quantum fluctuations as originally conjectured by Penrose \cite{Penrose}. The entanglement entropy of such a state scales as the area of the boundary of the region once a ultraviolet cutoff is introduced. The pre-inflationary scenario proposed here consists in a quantum BKL phase: the universe started in a low entanglement entropy state, with an entropy much lower than the one conjectured in Penrose's Weyl curvature hypothesis. In this phase the entanglement entropy grows with the scale factor until it reaches an equilibrium semiclassical state, in which a Bekenstein-Hawking like area law $S_R=\text{Area}(\partial R)/4G$ is expected for the entanglement entropy of a semiclassical region of space \cite{Bianchi:2012ev,VanRaamsdonk:2010pw}. Afterwards, the entanglement entropy of a comoving region scales as $a^2$, and the entanglement time then corresponds to the choice of the squared scale factor as relational time. This scenario therefore provides a mechanism for the production of the correlations present in the state at the beginning of inflation and imprinted today in the statistical fluctuations of the cosmic microwave background, as well as an entropic arrow of time applicable even in the quantum gravity regime where classical geometric notions may not be well defined.

\bigskip

\textbf{Acknowledgments}. We thank Abhay Ashtekar, Brajesh Gupt and Wolfgang Wieland for valuable discussions. This work is supported by the NSF grant PHY-1404204. NY acknowledges support from CNPq, Brazil.

\bigskip

%\begin{acknowledgments}
%We thank Abhay Ashtekar, Brajesh Gupt and Wolfgang Wieland for valuable discussions. The work of EB is supported by the NSF grant PHY-1404204. NY acknowledges support from CNPq, Brazil.
%\end{acknowledgments}

%
%\begin{equation}
%Q(a)=\left(
%\begin{array}{cc}
% 3\log \frac{a}{a_0} &  0   \\[1em]
%0  &   \frac{1}{2}(a^6-a_0^6)\, m^2-\frac{3}{4}(a^4-a_0^4)\, \Delta   
%\end{array}
%\right)
%\label{eq:}
%\end{equation}

\providecommand{\href}[2]{#2}\begingroup\raggedright\endgroup

\end{document}